\documentclass[preprint,aps,showpacs,nofootinbib]{revtex4}

\usepackage{epsfig}

\def\bea{\begin{eqnarray}}
\def\eea{\end{eqnarray}}
\def\bec{\begin{center}}
\def\ec{\end{center}}

\def\beq{\begin{equation}}
\def\eeq{\end{equation}}

\begin{document}
\draft
\tighten
\title{\large \bf Dynamical gauge coupling unification from moduli stabilization}
\author{
Kiwoon Choi\footnote{kchoi@hep.kaist.ac.kr}}
\address{
Department of Physics\\ Korea Advanced Institute of Science and
Technology\\ Daejeon 305-701, Korea}
\begin{abstract}
In $D$-brane models, different part of the 4-dimensional gauge
group might originate from $D$-branes wrapping different cycles in
the internal space, and then the standard model gauge couplings at
the compactification scale are determined by different
cycle-volume moduli. We point out that those cycle-volume moduli
can naturally have universal vacuum expectation values up to small
deviations suppressed by $1/8\pi^2$ if they are stabilized by
KKLT-type non-perturbative superpotential.  This dynamical
unification of gauge couplings is independent of the detailed form
of the moduli K\"ahler potential, but relies crucially on the
existence of low energy supersymmetry.  If supersymmetry is broken
by an uplifting brane as in KKLT compactification, again
independently of the detailed form of the moduli K\"ahler
potential, the  moduli-mediated gaugino masses at the
compactification scale are universal also, and are comparable to
the anomaly-mediated gaugino masses. As a result, both the gauge
coupling unification at high energy scale and the mirage mediation
pattern of soft supersymmetry breaking masses are achieved
naturally even when the different sets of the
standard model gauge
bosons originate from $D$-branes wrapping different cycles in the
internal space.
\end{abstract}
\maketitle

Low energy supersymmetry (SUSY) is one of the primary candidates
for physics beyond the standard model (SM) at TeV scale
\cite{Nilles:1983ge}. In addition to solving the hierarchy problem
between the weak scale and the GUT/Planck scale, the minimal
realization of low energy SUSY leads to a successful unification
of the $SU(3)_c\times SU(2)_W\times U(1)_Y$ gauge couplings at
$M_{\rm GUT} \sim 2 \times 10^{16}$ GeV \cite{unification}. There
are several different scenarios which would ensure the gauge
coupling unification at high energy scale in compactified string
theory. One possibility is that $G_{\rm SM}=SU(3)_c\times
SU(2)_W\times U(1)_Y$ is embedded into a simple gauge group
$G_{\rm GUT}$ at a scale around or below the compactification
scale. This would guarantees that the standard model couplings are
unified at the scale where $G_{\rm GUT}$ is broken down to $G_{\rm
SM}$. Even in the absence of such embedding, compactified string
theories with low energy SUSY offer an elegant mechanism to ensure
the gauge coupling unification. In many cases, the 4D gauge
coupling constants are determined by a single dilaton or modulus
up to small loop threshold corrections. Then the gauge kinetic
functions of $G_{\rm SM}$ are given by $f_a=k_aS$ ($a=3,2,1$)
where $k_a$ are {\it quantized} numbers which can be chosen to be
integers under a proper normalization of the dilaton superfield
$S$.
 In the framework of 4D effective theory,
the quantization of $k_a$ can be understood as the periodicity
condition of the axion field ${\rm Im}(S)$. If the integer
coefficients $k_a$ are chosen as $k_3=k_2=k_1$ (Here we are using
the $U(1)_Y$ hypercharge convention for which $g_1^2=g_2^2$.), the
standard model gauge couplings are unified at the compactification
scale $M_{\rm com}$ as desired, \bea \frac{1}{g_a^2(M_{\rm
com})}\equiv{\rm Re}(f_a)=k_a{\rm Re}(S),\eea and this is what
happens in typical heterotic string compactifications.

Recent development in string theory suggests that our world might
be described by $D$-brane models realized within type IIB or IIA
string theory \cite{shiu,gkp,kklt}. This set-up is particularly
interesting as it provides a framework of stabilizing all moduli,
leading to a landscape of string vacua which contains a
phenomenologically viable nearly flat de-Sitter (dS) vacuum with
low energy SUSY \cite{gkp,kklt}. It might be possible that the
entire $G_{\rm SM}$ is obtained from a stack of $D3$-branes or
$Dp$-branes ($p>3$) wrapping a common $(p-3)$ cycle in the
internal space. In such case, the gauge kinetic functions of
$G_{\rm SM}$ are determined by a single dilaton $S$ or modulus $T$
with integer coefficients: \bea f_a=k_a S \quad\mbox{or}\quad k_aT
\quad (a=3,2,1), \eea leading to the gauge coupling unification at
the string or compactification scale again under the choice
$k_1=k_2=k_3$. However such set-up typically suffers from the
difficulty of stabilizing open string moduli such as the
$D3$-brane position moduli and/or the Wilson line moduli for
symmetry breaking. In this regard, more interesting possibility is
that the different part of $G_{\rm SM}$ originates from
$Dp$-branes wrapping different cycles in the internal space. In
such models, the gauge kinetic functions of $G_{\rm SM}$ are given
by \bea \label{visiblegauge} f_a=k_aT_a\equiv k_a\left[\,
\frac{e^{-\phi}}{(2\pi)^{(p-2)}(\alpha^\prime)^{(p-3)/2}}V_a+iC_a\,\right],\eea
where $\phi$ is the string dilaton, $V_a$ denotes the volume of
the $a$-th $(p-3)$-cycle wrapped by the $Dp$-branes of the $a$-th
gauge group,  $C_a$ is the RR axion partner of $V_a$, and $k_a$ is
the number of windings. In order to realize the gauge coupling
unification in such case, one needs to adjust the dynamics of
moduli stabilization to make the vacuum expectation values of the
cycle-volumes $V_a$ ($a=3,2,1$) to be equal up to a relative
accuracy of ${\cal O}(1/8\pi^2)$, which might require non-trivial
fine-tunings of the parameters governing the stabilization of
$V_a$.

In this paper, we point out that the necessary dynamical gauge
coupling unification can be naturally achieved  if $T_a$ are
stabilized by KKLT-type non-perturbative superpotential
\cite{kklt}. Interestingly, the existence of low energy SUSY, i.e.
a large hierarchy between the Planck scale and the gravitino mass,
is crucial for this dynamical gauge coupling unification. It is
noted also that if SUSY is broken by an uplifting brane as in the
KKLT compactification of type IIB string theory \cite{kklt}, the
SUSY-breaking $F$-components of $T_a$, more precisely
$F^a/(T_a+T_a^*)$, also have universal vacuum expectation values
up to small deviations suppressed by $1/8\pi^2$ independently of
the moduli K\"ahler potential. This results in universal
moduli-mediated gaugino masses at the compactification scale,
which are comparable to the anomaly-mediated gaugino masses,
thereby yielding the mirage mediation pattern of low energy
superparticle masses which was discussed in
\cite{choi1,choi2,endo}. Furthermore, the axionic shift symmetries
for the RR-axions ${\rm Im}(T_a)$ assure that soft terms preserve
CP \cite{choi4} in this case of multi-messenger moduli
also\footnote{This is true for the gaugino masses and trilinear
scalar couplings, while we need additional mechanism to avoid a
dangerous phase of the Higgs $B$-parameter. However if the Higgs
sector of the model is given by the next-to-minimal supersymmetric
standard model (NMSSM), the axionic shift symmetries can assure
that all soft terms preserve CP as was stressed in \cite{choi2}.}.
Also the underlying geometric set-up for matter fields suggests
that it is rather plausible that moduli-mediated soft terms
preserve flavors. Thus both the gauge coupling unification at high
energy scale and the flavor and CP conserving mirage-mediation
pattern of low energy soft terms \cite{choi1,choi2} are achieved
naturally even when the standard model gauge bosons originate from
$Dp$ ($p>3$) branes wrapping different $(p-3)$-cycles in the
internal space.

Our discussion of dynamical gauge coupling unification  relies on
some characteristic features of KKLT-type compactification
\cite{kklt}. Let us thus consider the KKLT compactification of
type IIB string theory on Calabi-Yau (CY) orientifold as a
concrete example. Following \cite{gkp}, we consider a set-up in
which the string dilaton and all complex structure moduli are
stabilized by RR and NS-NS 3-form fluxes with heavy masses. We
then assume that all visible sector gauge groups originate from
$D7$ branes wrapping the 4-cycles $\Sigma_a$ of CY orientifold.
The visible sector gauge kinetic functions in such set-up are
given by (\ref{visiblegauge}) with $p=7$, where now $T_a$ denote
the K\"ahler moduli of CY orientifold and $C_a$ are the RR 4-form
axions. Following KKLT \cite{kklt}, let us assume that $T_a$ are
stabilized by non-perturbative effects such as $D3$-brane
instantons wrapping $\Sigma_a$ or hidden gaugino condensations on
$D7$ branes wrapping $\Sigma_a$. After the string dilaton and
complex structure moduli are integrated out, these
non-perturbative effects yield the following form of the effective
superpotential: \bea \label{nonperturbative} W_{\rm
eff}=w_0+\sum_a A_ae^{-8\pi^2r_aT_a} \eea where $w_0$ and $A_a$
are effective parameters depending on the vacuum expectation
values of heavy complex structure moduli, while $r_a$ are (either
topological or group theoretical) {\it rational} numbers. Without
loss of generality, one can always make $\omega_0$ and $A_a$ real
by an appropriate $U(1)_R$ transformation and also the axionic
shift transformations: \bea \label{axionshift} T_a\rightarrow
T_a+\mbox{imaginary constant}. \eea In the following, we will use
such field basis in which $\omega_0$ and $A_a$ are real
parameters. Note that yet this does not mean that the vacuum
expectation value of $W_{\rm eff}$ is real.

For $D3$-brane instantons and $D7$-brane gaugino condensations,
the corresponding coefficients $A_a$ are generically of order
unity. On the other hand, $\omega_0$ is required to be small to
get low energy SUSY since it determines the gravitino mass as
$\omega_0\sim m_{3/2}$ in the unit with the 4D Planck scale
$M_{Pl}=1$. For instance, one needs $\omega_0\sim 10^{-14}$ to get
$m_{3/2}\sim 10$ TeV  which would be necessary for low energy SUSY
in our case. Such  hierarchically small value of $\omega_0$ might
be able to be obtained for a particular set of RR and NS-NS flux
configurations in the landscape of type IIB flux vacua
\cite{douglas}. Alternatively, there might be an $R$-symmetry
which is preserved by fluxes,  but is broken only by
non-perturbative effects. In such case, hierarchically small
$\omega_0$ can be generated naturally  by gaugino condensation on
$D3$ brane, yielding $\omega_0\sim e^{-8\pi^2 S/N}$ which becomes
an exponentially small constant after the type IIB dilaton $S$ is
integrated out. Here we simply assume that $w_0$ has a correct
value for low energy SUSY, e.g. $w_0\sim m_{3/2}\sim 10^{-14}$ in
the unit with $M_{Pl}=1$, without specifying its origin. In fact,
this large hierarchy between $m_{3/2}$ and $M_{Pl}$ is a key
element of the dynamical gauge coupling unification which will be
discussed below.

As for the quantized coefficients $r_a$, if the non-perturbative
superpotential of $T_a$ is due to $D3$ brane instanton, one finds
$r_a=1$ \cite{polchinski}. Another possible origin of the
non-perturbative superpotential is hidden gaugino condensation
\cite{gauginocon}. Suppose that $D7$ branes wrapping $\Sigma_a$
provides a supersymmetric hidden $SU(N_a)$ gauge theory with the
gauge kinetic function $f_H=k^{(H)}_aT_a$ in addition to the
$a$-th visible sector gauge group with the gauge kinetic function
$f_a=k_aT_a$, where again we choose the normalization of $T_a$ for
which $k_a$ and $k^{(H)}_a$ are integers\footnote{In 4D effective
theory, one might redefine $T_a$ in a way for which either $k_a=1$
or $r_a=1$. Here we use the definition of $T_a$ given in
(\ref{visiblegauge}), for which $k_a$ are  integers and $r_a$ are
rational numbers.}. Then the gaugino condensation of $SU(N_a)$
gives $r_a=k^{(H)}_a/N_a$. In fact, string theoretic determination
of $k_a$ and $r_a$ is not our major concern here as our subsequent
discussion will rely {\it only} on that both $k_a$ and $r_a$ are
(topological or group theoretical) {\it rational} numbers, which
can be ensured by the periodicity of the axion field ${\rm
Im}(T_a)$ within the framework of 4D effective theory.

Let us examine the 4D gauge couplings in
 the model with the visible sector gauge kinetic function
(\ref{visiblegauge}) and the moduli superpotential
(\ref{nonperturbative}).
 As is well known, the moduli
superpotential (\ref{nonperturbative}) stabilizes $T_a$ at the
supersymmetric AdS vacuum \cite{kklt} satisfying \bea
\label{susyads} D_aW_{\rm eff}=\partial_aW_{\rm
eff}+\partial_aK_0W_{\rm eff}=0, \eea where $K_0$ denotes the
moduli K\"ahler potential. The resulting vacuum expectation values
of ${\rm Re}(T_a)$ are given by \bea \label{vacuumvalue}r_a\langle
{\rm Re}(T_a)\rangle
=\frac{1}{8\pi^2}\ln\left(\frac{8\pi^2M_{Pl}}{|m_{3/2}|}\right)
+\frac{1}{8\pi^2}\ln\left(\frac{e^{K_0/2}r_aA_a}{|\partial_a
K|}\right), \eea where $m_{3/2}=e^{K_0/2}W_{\rm eff}$. As
$A_a,r_a,
\partial_aK_0$ and $e^{K_0/2}$ are generically of order unity, while
$M_{Pl}/m_{3/2}\sim 10^{14}$ for $m_{3/2}\sim 10$ TeV, the above
result shows that even when $A_a, r_a$ and $\partial_a K$ are {\it
not} universal,  $r_a{\rm Re}(T_a)$ have universal vacuum
expectation values ($\sim1/2$) up to small deviations suppressed
by $1/8\pi^2$. The resulting 4D gauge couplings at the
compactification scale $M_{\rm com}$ are given by \bea
\frac{1}{g_a^2(M_{\rm com})}
=\frac{k_a}{2r_a}\left[\frac{\ln(8\pi^2M_{Pl}/|m_{3/2}|)}{4\pi^2}+{\cal
O}\left(\frac{1}{8\pi^2}\right)\right]=\frac{k_a}{2r_a}\left[1+{\cal
O}\left(\frac{1}{8\pi^2}\right)\right], \eea where we have used
$\ln(8\pi^2M_{Pl}/|m_{3/2}|)\simeq 4\pi^2$ for the gravitino mass
$m_{3/2}\sim 10$ TeV. Thus if one chooses the (topological or
group theoretical) rational numbers $k_a$ and $r_a$ as \bea
\label{unification} k_a/r_a=4, \eea one achieves the successful
gauge coupling unification at $M_{\rm com}$ with the correct
unified coupling constant $g_{GUT}^2\simeq 2$ even though
different set of gauge bosons originate from $D7$ branes wrapping
different 4-cycles. Note that the scheme is totally independent of
the form of the moduli K\"ahler potential, while the existence of
large hierarchy $M_{Pl}/m_{3/2}\sim 10^{14}$ which is necessary
for low energy SUSY is crucial for this dynamical gauge coupling
unification.

The universal vacuum values of $r_a{\rm Re}(T_a)$ in
(\ref{vacuumvalue}) correspond to a supersymmetric AdS vacuum. To
obtain phenomenologically viable dS (or Minkowski) vacuum, one
needs to introduce a SUSY-breaking brane into the above moduli
stabilization set-up, providing an additional moduli potential
lifting the AdS vacuum of (\ref{vacuumvalue}) to a dS vacuum
\cite{kklt}. Then one should make sure that the universality of
the moduli vacuum values is not spoiled by the vacuum shift
induced by the uplifting potential. As we will see, the vacuum
shifts due to the uplifting potential are small enough not to
destabilize the dynamical gauge coupling unification set by
$W_{\rm eff}$:
 \bea
 \label{smallshift}
\delta{\rm Re}(T_a)= {\cal
O}\left(\frac{m_{3/2}^2}{m_{T_a}^2}\right)={\cal
O}\left(\frac{1}{(8\pi^2)^2}\right). \eea

Let us now discuss the SUSY-breaking induced by an uplifting brane
in the above scenario of dynamical gauge coupling unification.
 The uplifting brane might be an anti-$D3$ brane as in the
original KKLT  proposal \cite{kklt} or a different kind of brane
on which SUSY is spontaneoulsy broken \cite{nilles,hebecker1}. In
fact, our whole discussion is independent of the detailed property
of the uplifting brane because of the following reason. One of the
generic features of IIB flux compactification on CY orientifold is
the presence of warped throat. In the presence of such warped
throat, the SUSY-breaking brane will be stabilized at the end of
(maximally) warped throat in order to minimize its contribution to
the 4D effective potential \cite{kklt}. As was discussed in
\cite{choi1}, low energy consequences of such {\it red-shifted}
SUSY-breaking brane can be described by a {\it single} $D$-type
spurion operator whose coefficient is uniquely fixed by the
condition of nearly vanishing 4D cosmological constant. More
explicitly, independently of the detailed property of the
red-shifted uplifting brane, low energy physics of the visible
fields and light moduli can be described by the following form of
4D effective action:
 \bea
 \label{superspace}
 \int d^4\theta \left[-3CC^*e^{-K/3}
-e^{4A}C^2C^{*2}{\cal P}_{\rm
lift}\theta^2\bar{\theta}^2\right]+\left(\int d^2\theta \left[
\frac{1}{4}f_aW^{a\alpha}W^a_\alpha+C^3W\right]+{\rm h.c.}
\right), \eea where $C=C_0+F^C\theta^2$ is the chiral compensator
superfield,  $e^A$ is the energy red-shift factor on the
world-volume of SUSY-breaking brane \cite{rs}, $K$ and $W$ are the
conventional K\"ahler potential and superpotential which can be
expanded in powers of the visible matter superfields
$Q^i$: \bea K&=&K_0(T_a,T_a^*)+Z_i(T_a,T_a^*)Q^{i*}Q^i, \nonumber \\
W&=&W_{\rm eff}(T_a)+\frac{1}{6}\lambda_{ijk}(T_a) Q^iQ^jQ^k.\eea

In the Einstein frame, the moduli potential from the superspace
lagrangian density (\ref{superspace}) takes the form:
 \bea V_{\rm TOT}
=e^{K_0}\left[K^{a\bar{b}}D_aW_{\rm eff}(D_bW_{\rm
eff})^*-3|W_{\rm eff}|^2\right]+V_{\rm lift}, \eea where \bea
\label{uplift1} V_{\rm lift}= e^{4A}e^{2K_0/3}{\cal P}_{\rm
lift}.\eea The superspace lagrangian density (\ref{superspace})
also determines the auxiliary components of $C$ and $T_a$ as
 \bea \frac{F^C}{C_0}&=& m_{3/2}^*+\frac{1}{3}F^a\partial_aK, \nonumber
\\
F^a&=& -e^{K_0/2}K^{a\bar{b}}(D_bW_{\rm eff})^* .\eea To analyze
SUSY-breaking, one needs first to compute the vacuum expectation
values of $F^a$ determined by $V_{\rm TOT}$. To this end, let us
expand $V_{\rm TOT}$ and $F^a$ around the supersymmetric
configuration satisfying (\ref{susyads}). We then find \bea V_{\rm
TOT}&=& -3|m_{3/2}(\vec{T}_0)|^2+V_{\rm lift}(\vec{T}_0)
+\partial_aV_{\rm lift}(\vec{T}_0)\delta
T_a+\partial_{\bar{a}}V_{\rm lift}(\vec{T}_0)\delta T^*_a
\nonumber \\
&&+ m^2_{a\bar{b}}(\vec{T}_0)\delta T_a\delta
T_b^*+\frac{1}{2}\left( \mu^2_{ab}(\vec{T}_0)\delta T_a\delta
T_b+{\rm h.c.}\right)+{\cal O}((\delta T)^3), \nonumber \\
F^a&=& -K^{a\bar{b}}v^*_{\bar{b}\bar{c}}\delta
T^*_c-m_{3/2}^*\delta T_a+{\cal O}((\delta T)^2), \eea where
$T_a=T_{0a}+\delta T_a$ for the supersymmetric moduli
configuration $T_{0a}$ satisfying (\ref{susyads}), and \bea
m^2_{a\bar{b}}&=&
K^{c\bar{d}}v_{ac}v^*_{\bar{b}\bar{d}}-2|m_{3/2}|^2\partial_a\partial_{\bar{b}}K_0,
\nonumber
\\\mu^2_{ab}&=&-m_{3/2}^*v_{ab}, \eea for
\bea v_{ab} =e^{K_0/2}\left[\partial_a\partial_b W_{\rm
eff}+(\partial_a\partial_b K_0-\partial_a K_0\partial_b K_0)W_{\rm
eff}\right]. \eea For the superpotential (\ref{nonperturbative}),
we have \bea
\partial_a\partial_b W_{\rm eff}&=&-8\pi^2
r_a\partial_aW_{\rm eff}\delta_{ab}= 8\pi^2 r_a\partial_a
K_0W_{\rm eff}\delta_{ab}, \eea where we have used the SUSY
condition (\ref{susyads}) for the last expression.

Since $r_a{\rm Re}(T_a)\simeq 1/2$ as derived in
(\ref{vacuumvalue}), generically $\partial_a\partial_b K_0\sim
\partial_aK_0\partial_bK_0\sim r_a\partial_a K_0$, and thus
\bea \label{result1} v_{ab}&=&8\pi^2r_a\partial_a K_0
m_{3/2}\delta_{ab}\left(1+{\cal
O}\left(\frac{1}{8\pi^2}\right)\right)
={\cal O}(8\pi^2 m_{3/2}), \nonumber \\
m^2_{a\bar{b}}&=&(8\pi^2)^2K^{a\bar{b}}r_ar_b\partial_aK_0\partial_{\bar{b}}K_0|m_{3/2}|^2
\left(1+{\cal O}\left(\frac{1}{8\pi^2}\right)\right)={\cal
O}((8\pi^2 m_{3/2})^2), \nonumber \\
F^a&=&-8\pi^2K^{a\bar{b}}r_b\partial_{\bar{b}}K_0m_{3/2}^*\delta
T^*_b\left(1+{\cal O}\left(\frac{1}{8\pi^2}\right)\right)={\cal
O}(8\pi^2 m_{3/2}\delta T_a).
 \eea Then the equation of motion $\partial_a V_{\rm TOT}=0$
 leads to \bea (8\pi^2)^2|m_{3/2}|^2\sum_b K^{a\bar{b}}r_ar_b\partial_{\bar{b}}K_0\delta T^*_b=
-\frac{\partial_aV_{\rm lift}}{\partial_a K_0} \left(1+{\cal
O}\left(\frac{1}{8\pi^2}\right)\right).\eea Combining this with
the condition of vanishing cosmological constant, we find \bea
\label{result2} V_{\rm
lift}(\vec{T}_0)&=&3|m_{3/2}(\vec{T_0})|^2\left(1+{\cal
O}(\frac{1}{(8\pi^2)^2})\right), \nonumber \\
\sum_b K^{a\bar{b}}r_ar_b\partial_{\bar{b}}K_0\delta
T^*_b&=&-\frac{3}{(8\pi^2)^2}\frac{\partial_a\ln(V_{\rm
lift})}{\partial_a K_0} \left(1+{\cal
O}\left(\frac{1}{8\pi^2}\right)\right), \eea which shows that
indeed \bea \delta T_a={\cal
O}\left(\frac{1}{(8\pi^2)^2}\right)\eea as anticipated in
(\ref{smallshift}). Applying (\ref{result2}) to $F^a$ in
(\ref{result1}), we finally find \bea
\label{result3}\frac{F^a}{T_a+T_a^*}&=&
\frac{m_{3/2}^*}{8\pi^2}\frac{1}{r_a(T_a+T_a^*)}\frac{3\partial_a\ln(V_{\rm
lift})}{\partial_a K}\left(1+{\cal
O}\left(\frac{1}{8\pi^2}\right)\right) \nonumber \\
&=&\frac{m_{3/2}^*}{\ln(M_{Pl}/|m_{3/2}|)}
\frac{3\partial_a\ln(V_{\rm lift})}{2\partial_a K}\left(1+{\cal
O}\left(\frac{1}{8\pi^2}\right)\right), \eea where we have used
the moduli vacuum values (\ref{vacuumvalue}) for the last
expression.

The above result shows that $F^a/(T_a+T_a^*)$ are universal if
$\partial_a \ln(V_{\rm lift})/\partial_a K$ are universal. In
fact, $\partial_a\ln(V_{\rm lift})/\partial_a K_0$ are universal
independently of the form of $K_0$ in the most interesting case
that  SUSY-breaking brane is stabilized at the end of warped
throat \cite{kklt}. In this case, $T_a$ are sequestered from the
SUSY-breaking brane, which means that ${\cal P}_{\rm lift}$ is
independent of $T_a$ \cite{choi1,hebecker}. One then has $V_{\rm
lift}=De^{2K_0/3}$ for a constant $D$ of ${\cal
O}(m_{3/2}^2M_{Pl}^2)$, yielding \bea \frac{F^a}{T_a+T_a^*}=
\frac{m_{3/2}^*}{\ln(M_{Pl}/|m_{3/2}|)}\left(1+{\cal
O}\left(\frac{1}{8\pi^2}\right)\right). \eea One might consider a
more general possibility that the uplifting function ${\cal
P}_{\rm lift}$ depends on $T_a$, but only through $K_0$, i.e. \bea
\label{uni} {\cal P}_{\rm lift}={\cal P}_{\rm
lift}[K_0(T_a,T_a^*)]. \eea In this more general case,
$\partial_a\ln(V_{\rm lift})/\partial_a K_0$ are still universal,
yielding \bea \label{fterm} \frac{F^a}{T_a+T_a^*}=
\frac{m_{3/2}^*}{\ln(M_{Pl}/|m_{3/2}|)}
\left(1+\frac{3}{2}\frac{d\ln({\cal P}_{\rm
lift})}{dK_0}\right)\left(1+{\cal
O}\left(\frac{1}{8\pi^2}\right)\right). \eea Note that both the
magnitudes and phases of $F^a/(T_a+T_a^*)$ are universal
independently of the form of $K_0$ and also of the values of the
quantized numbers $k_a$ and $r_a$. This universality is valid even
when the condition (\ref{unification}) for the gauge coupling
unification is not satisfied. What is required is only the
condition (\ref{uni}) which is satisfied for the case that
SUSY-breaking brane is stabilized at the end of a warped throat.

So far, we have shown that $r_a{\rm Re}(T_a)$ and
$F^a/(T_a+T_a^*)$ have universal vacuum expectation values up to
small deviations suppressed by $1/8\pi^2$. With universal
$r_a\langle {\rm Re}(T_a)\rangle$, we can realize a dynamical
unification of gauge couplings even when the standard model gauge
bosons originate from $D$ branes wrapping different cycles in the
internal space.   Let us now consider possible implications of the
universality of $F^a/(T_a+T_a^*)$ for soft SUSY-breaking terms of
visible fields, in particular for the SUSY CP and flavor problems.
Since we have relatively many (equally important) sources of
SUSY-breaking, i.e. $F^a$ ($a=1,2,3,...$) and $F^C$, one might
worry that the resulting soft terms can cause dangerous SUSY CP
and/or flavor violation. However it turns out that the set-up can
preserve CP and flavor in a rather natural way.

The soft SUSY-breaking terms of canonically normalized visible
fields can be written as
\begin{eqnarray}
{\cal L}_{\rm
soft}&=&-\frac{1}{2}M_a\lambda^a\lambda^a-\frac{1}{2}m_i^2|\tilde{Q}^i|^2
-\frac{1}{6}A_{ijk}y_{ijk}\tilde{Q}^i\tilde{Q}^j\tilde{Q}^k+{\rm
h.c.},
\end{eqnarray}
where $\lambda^a$ are gauginos, $\tilde{Q}^i$ is the scalar
component of the matter superfield $Q^i$, and $y_{ijk}$ are the
canonically normalized Yukawa couplings:
$y_{ijk}={\lambda}_{ijk}/\sqrt{e^{-{K}_0}Z_iZ_jZ_k}$. The most
interesting feature of the SUSY-breaking by red-shifted uplifting
brane is that the induced $F$-components of $T_a$ are of ${\cal
O}(m_{3/2}/8\pi^2)$ as can be seen in (\ref{result3}). As a
result, the tree-level moduli-mediated soft masses of ${\cal
O}(F^a/T_a)$ are comparable to the 1-loop anomaly-mediated soft
masses of ${\cal O}(m_{3/2}/8\pi^2)$ \cite{amsb}, leading to the
mirage mediation pattern of superparticle masses at low energy
scale \cite{choi1,choi2,endo}. Note that the 1-loop
anomaly-mediation does not depend on the UV physics above the
compactification scale as it is determined by the infrared
structure of 4D effective theory. On the other hand, 1-loop
threshold corrections to the moduli mediation at the
compactification scale are generically sensitive to the physics
above the compactification scale. Such UV-sensitive corrections to
the soft masses are of ${\cal O}(F^a/8\pi^2T_a)$, thus can be
safely ignored in our computation of soft terms at the
compactification scale. Then from the superspace action
(\ref{superspace}), one finds that the soft masses renormalized at
 $M_{\rm com}$ are given by\footnote{There might be model-dependent extra contribution to
 $m_i^2$ mediated by a light vector multiplet in the warped throat
 \cite{choi5,hebecker1}, however it does not affect our discussion in this paper.}
\begin{eqnarray}
\label{soft1} M_a(M_{\rm com})&=& \sum_b
F^b\partial_b\ln\left({\rm Re}(f_a)\right)
+\frac{b_ag_a^2}{8\pi^2}\frac{F^C}{C_0},
\nonumber \\
A_{ijk}(M_{\rm com})&=& -\sum_b
F^b\partial_b\ln\left(\frac{{\lambda}_{ijk}}{e^{-{
K}_0}Z_iZ_jZ_k}\right)-
\frac{1}{16\pi^2}(\gamma_i+\gamma_j+\gamma_k)\frac{F^C}{C_0},
\nonumber \\
m_i^2(M_{\rm com})&=& \frac{2}{3}V_{\rm TOT} -\sum_{b,c}F^b
F^{c*}\partial_b\partial_{\bar{c}}\ln \left(e^{-{
K}_0/3}Z_i\right)\nonumber
\\ &&
-\frac{1}{32\pi^2}\frac{d\gamma_i}{d\ln\mu}\left|\frac{F^C}{C_0}\right|^2
 + \frac{1}{16\pi^2}\left( \sum_b\partial_{b}{\gamma}_i
F^b\left(\frac{F^C}{C_0}\right)^* +{\rm h.c.}\right),
\end{eqnarray}
where $b_a$ and $\gamma_i$ are the one-loop beta function
coefficients and the anomalous dimension of $Q^i$, respectively,
defined as ${dg_a}/{d\ln \mu}={b_ag_a^3}/{8\pi^2}$ and ${d\ln
Z_i}/{d\ln \mu}=\gamma_i/{8\pi^2}$.

For $f_a$ given by (\ref{visiblegauge}), the gaugino masses are
given by \bea \label{gauginomass}M_a(M_{\rm com})&=&
\frac{F^a}{(T_a+T_a^*)}+\frac{b_a}{8\pi^2}g_a^2m_{3/2}^* \nonumber
\\&=& \left(1+\alpha\,\frac{\ln(M_{Pl}/|m_{3/2}|)}{8\pi^2}b_ag_a^2(M_{\rm
com})\right)M_0, \eea where $M_0$ denotes the {\it universal}
moduli-mediated gaugino mass at $M_{\rm com}$, i.e.
 \bea M_0\equiv\frac{F^a}{T_a+T_a^*},
 \eea
 and $\alpha$ represents the moduli-mediation  to anomaly-mediation
 ratio\footnote{Clearly $\alpha=1$ is the most plausible value of the moduli-mediation to anomaly-mediation ratio
 as it is predicted by the simplest class of models
with a sequestered uplifting brane for which ${\cal P}_{\rm lift}$
is independent of $T_a$. Still it is possible to get a different
value of $\alpha$  while keeping ${\cal P}_{\rm lift}$ independent
of $T_a$ by generalizing for instance the gauge kinetic functions
as $f_a=k_aT_a+\mbox{constant}$, while keeping $W_{\rm eff}$
 unchanged \cite{tatsuo,choi5}}:
 \bea
 \alpha\equiv \frac{m_{3/2}^*}{M_0\ln(M_{Pl}/|m_{3/2}|)}=
1+ \frac{3}{2}\frac{d\ln({\cal P}_{\rm lift})}{dK_0}. \eea A
notable features of the above gaugino masses is that the
moduli-mediated gaugino masses at $M_{\rm com}$ are universal
(both the magnitudes and phases) independently of the form of
$K_0$, and also independently of the discrete numbers $k_a$ and
$r_a$, thus independently of whether the gauge coupling
unification condition (\ref{unification}) is satisfied. Another
interesting feature is that the universal moduli-mediated gaugino
mass $M_0$ has the same phase as the anomaly-mediated gaugino
masses, thus there is no CP-violating relative phase between
different gaugino masses, again independently of the form of $K_0$
and the values of $k_a$ and $r_a$. The subsequent renormalization
group evolution of (\ref{gauginomass}) leads to the mirage
mediation pattern of low energy gaugino masses: \bea
M_a(\mu)=M_0\left[1-\frac{1}{4\pi^2}b_ag_a^2(\mu)\ln\left(
\frac{M_{\rm
com}}{(M_{Pl}/|m_{3/2}|)^{\alpha/2}\mu}\right)\right], \eea with
the mirage messenger scale $M_{\rm mirage}=M_{\rm
com}/(M_{Pl}/|m_{3/2}|)^{\alpha/2}$. Thus even when the  standard
model gauge bosons originate from $D7$-branes wrapping different
4-cycles in CY orientifold, a CP-conserving mirage mediation
pattern of low energy gaugino masses is obtained {\it
independently of} the form of the moduli K\"ahler potential and
also of the discrete parameters $k_a$ and $r_a$ which determine
the gauge kinetic functions and the non-perturbative
superpotential.

In fact, the (approximate) CP-invariance of the full soft terms
can be ensured by adopting a string theoretic property of the RR
axions ${\rm Im}(T_a)$, i.e. the property that the axionic shift
symmetry under the transformation (\ref{axionshift}) is broken
only by non-perturbative effects \cite{choi4}. For the moduli
vacuum values (\ref{vacuumvalue}),  those non-perturbative
corrections to the moduli K\"ahler potential $K_0$, the matter
K\"ahler metric $Z_i$ and the holomorphic Yukawa couplings
$\lambda_{ijk}$ are negligible compared to the leading order
terms, thus $K_0$, $Z_i$ and $\lambda_{ijk}$ are all invariant
under $$T_a\rightarrow T_a+\mbox{imaginary constant}.$$ This means
that $K_0$ and $Z_i$ depend only on the combinations $T_a+T_a^*$,
and $\lambda_{ijk}$ are $T_a$-independent constants: \bea
K_0&=&K_0(T_a+T_a^*),
\nonumber \\
 Z_i&=&Z_i(T_a+T_a^*),
 \nonumber \\
 \lambda_{ijk}&=&\mbox{constants}.
\eea It is rather obvious that for the above forms of $K_0$, $Z_i$
and $\lambda_{ijk}$, the soft parameters of (\ref{soft1}) preserve
$CP$. Note that $F^a$ and $F^C/C_0=m_{3/2}^*+F^a\partial_aK_0/3$
have a common phase as is shown in (\ref{fterm}), and this common
phase can be rotated away by an appropriate field redefinition. As
for the flavor issue, a crucial point might be that matter
superfields with the same gauge quantum numbers originate from the
intersection of the same pair of 4-cycles. It is then expected
that the K\"ahler metrics of the matter fields with same gauge
quantum numbers have a common dependence on the 4-cycle volume
moduli $T_a$\footnote{The matter K\"ahler metric $Z_i$ and the
holomorphic Yukawa couplings $\lambda_{ijk}$ generically have a
flavor non-universal dependence on the complex structure moduli,
which might be responsible for the hierarchical structure of the
canonical Yukawa couplings
$y_{ijk}=\lambda_{ijk}/\sqrt{e^{-K_0}Z_iZ_jZ_k}$. In our set-up,
the complex structure moduli $Z^\alpha$ are assumed to get heavy
masses $m_{Z^\alpha}\gg 8\pi^2 m_{3/2}$ from fluxes, thus their
contributions to soft masses are negligible as $F^{Z^\alpha}\sim
m_{3/2}^2/m_{Z^\alpha}\ll m_{3/2}/8\pi^2$.}, which would ensure
that the soft terms of (\ref{soft1}) mediated by $T_a$ and $C$
preserve flavor also.

To conclude, we pointed out that the gauge coupling unification at
high energy scales can be naturally achieved even when the
different sets of the standard model gauge bosons originate from
$Dp$-branes wrapping different $(p-3)$-cycles in the internal
space, if the gauge coupling moduli are stabilized by KKLT-type
superpotential. This dynamical gauge coupling unification  is
independent of the form of the moduli K\"ahler potential, but
relies crucially on the existence of low energy SUSY with which
$\ln(M_{Pl}/m_{3/2})\sim 4\pi^2$. We also examined the
SUSY-breaking due to an uplifting brane in the scenario of
dynamical gauge coupling unification. If the uplifting brane is
stabilized at the end of warped throat as in the KKLT
compactification of type IIB string theory, the resulting
moduli-mediated gaugino masses at the compactification scale are
universal independently of the form of the moduli K\"ahler
potential and also independently of the discrete parameters which
determine the gauge kinetic functions and the nonperturbative
superpotential. The universal moduli-mediated gaugino mass is
comparable to the anomaly-mediated gaugino masses, thereby
yielding the mirage mediation pattern of low energy superparticle
masses as was discussed before \cite{choi1,choi2,endo}. There are
relatively many (equally important) sources of SUSY-breaking, i.e.
the auxiliary components of the gauge coupling moduli $T_a$
($a=1,2,3,...$) and the chiral compensator superfield $C$, thus
one might worry about dangerous SUSY CP and/or flavor violations.
It turns out that soft terms naturally preserve CP as a
consequence of the axionic shift symmetries for the RR axion
components of the gauge coupling moduli. As for the SUSY flavor
conservation, since the matter superfields with same gauge quantum
number originate from the intersection of same pair of cycles, it
is expected that their K\a"hler metric has a flavor-universal
dependence on the cycle volume moduli $T_a$, leading to
flavor-conserving soft terms. Thus both the gauge coupling
unification at high energy scale and the flavor and CP conserving
mirage mediation pattern of low energy soft terms can be naturally
obtained even when the standard model gauge bosons originate from
$D$-branes wrapping different cycles in the internal space.

 \vspace{5mm}
\noindent{\large\bf Acknowledgments} \vspace{5mm}

I thank K. S. Choi, M. Cvetic, A. Hebecker, T. Kobayashi, H. P.
Nilles and A. Uranga for helpful discussions.  I also thank the
Galileo Galilei Institute for Theoretical Physics for the
hospitality and the INFN for partial support during the completion
of this work. This work is supported by the KRF Grant funded by
the Korean Government (KRF-2005-201-C00006), the KOSEF Grant
(KOSEF R01-2005-000-10404-0), and the Center for High Energy
Physics of Kyungpook National University.

\end{document}